\documentclass[fleqn,twoside]{article}
\usepackage{espcrc2}

\usepackage[
]{graphicx}
\usepackage{epstopdf}
\usepackage{bm}

\newcommand{\AmS}{{\protect\the\textfont2
  A\kern-.1667em\lower.5ex\hbox{M}\kern-.125emS}}

\title{\vspace{-5.0cm} 
\begin{flushright}
{\normalsize Talk presented at ``Lattice 2003'', July 15-19, 2003, Tsukuba, Japan}\\
\vspace{-0.2cm}
{\normalsize KEK-TH-915}\\
\vspace{-0.2cm}{\normalsize RBRC-334}\\
\end{flushright}
\vspace*{2.5cm}
Nucleon axial charge and structure functions with domain wall fermions}

\author{
Shigemi Ohta\address[KEK]{Institute of Particle and Nuclear Studies, KEK,  
Tsukuba, Ibaraki 305-0801, Japan}\address[RBRC]{RIKEN BNL Research Center, Brookhaven National Laboratory, Upton, NY 11973, USA} and
Kostas Orginos\addressmark[RBRC] for the RIKEN-BNL-Columbia-KEK Collaboration\thanks{Talk by SO.  We thank RIKEN, Brookhaven National Laboratory and the U.S.\ Department of Energy for providing the facilities essential for the completion of this work.}}
       
\begin{document}

\begin{abstract}
We report the current status of RBCK calculations on nucleon structure
with both quenched and unquenched lattice QCD.  The combination of
domain wall fermions and DBW2 gauge action works well for isovector
vector and axial charges, and moments of structure functions \(\langle
x \rangle_q\), \(d_1\), and \(\langle 1 \rangle_{\delta q}\).
\end{abstract}

\maketitle
 Domain wall fermions (DWF) \cite{DWF} have been successfully used by the
 RBC Collaboration for several applications \cite{RBC}.  
In the baryon sector we had reported that DWF with Wilson gauge action in the
quenched approximation can reproduce the first
negative-parity excited state of nucleon \cite{Sasaki:2001nf} which
defied theoretical attempts for a long time.
 Here we report the
current status of our calculations on the nucleon electroweak matrix
elements, namely the isovector vector and axial charges and some
moments of the structure functions, in both quenched and full-QCD,
using DWF and the DBW2 gauge action \cite{DBW2}.

Four form factors appear in neutron \(\beta\) decay: the vector and
induced tensor form factors from the vector current, and the
axial-vector and induced pseudo-scalar form factors from the
axial-vector current.  In the forward limit, the 4-momentum transfer
should be small because the mass difference of the neutron and proton
is only about 1.3 MeV.  This makes the limit \(q^2 \rightarrow 0\),
where the vector and axial-vector form factors dominate, a good
approximation.  Their values in this limit are called the vector and
axial charges of the nucleon: \(g_{_V}\) and \(g_{_A}\).

\begin{figure}
\begin{center}
\includegraphics[width=70mm,bb=0 0 547 247]{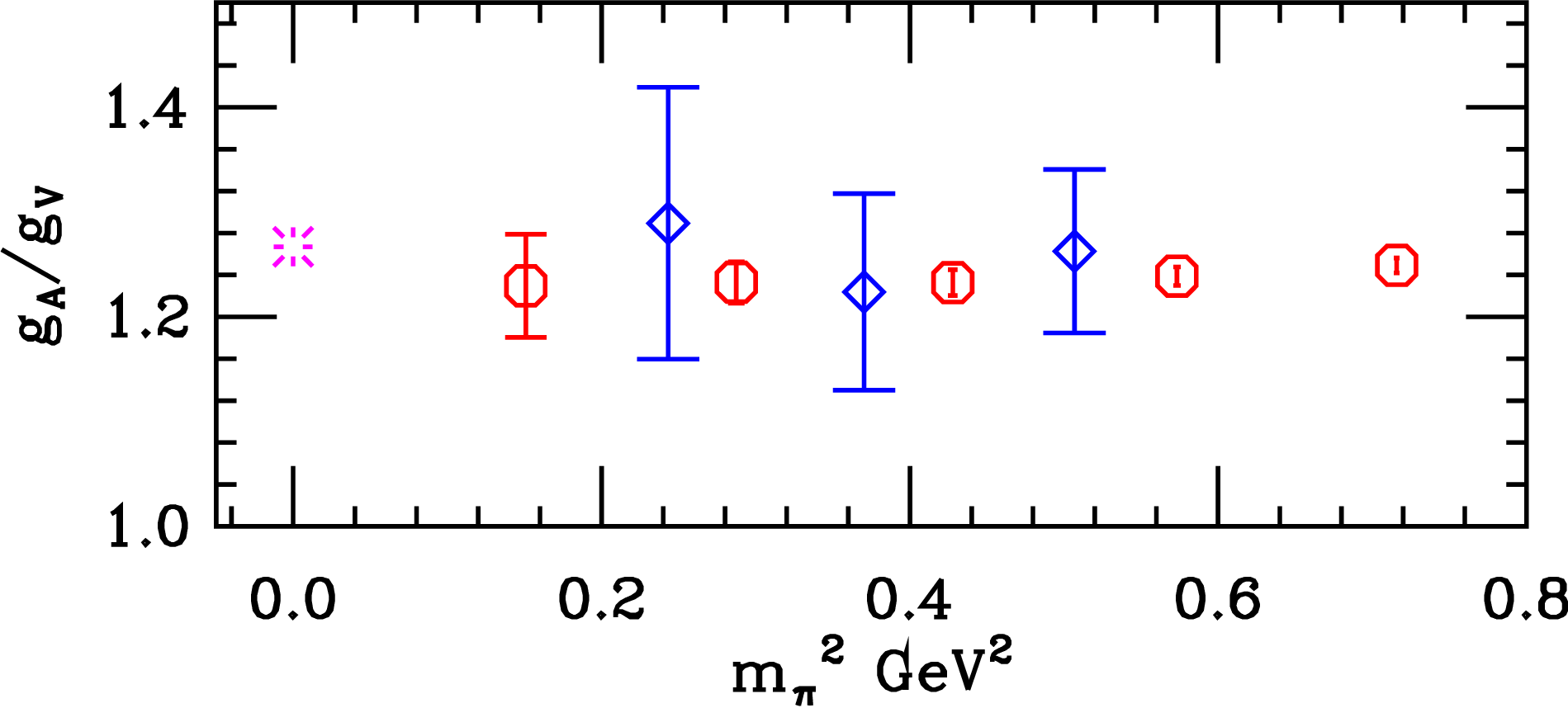}
\end{center}
\vspace{-1cm}
\caption{Quark mass dependence of axial to vector charge ratio, \(g_{_A}/g_{_V}\), quenched (circle) and full QCD (diamond).  Full-QCD results, albeit with small statistics, are consistent with quenched results:  No mass dependence is seen.}
\label{fig:mfdependence}
\vspace{-.5cm}
\end{figure}
Lattice numerical calculation of these charges is relatively simple:
prepare appropriate nucleon source and sink at rest, insert an
appropriate current operator in between, and identify a plateau which
gives a lattice estimate.  There already had been several preceding
works using Wilson or improved Wilson fermions \cite{Wilson}.  However
since these fermions schemes do not respect the chiral symmetry, the
lattice renormalizations of the relevant vector and axial currents are
not the same.  Worse, unwanted lattice artifact may result in
unphysical mixing of chirally distinct operators.  The combination of
DWF and DBW2 preserves chiral symmetry well enough so that we are free
of these difficulties: especially the relation \(Z_{_A} = Z_{_V}\) is
easily and well maintained, up to \(O(a^2)\), or a few \% at  \(a\approx 0.15\)
fm.  And in contrast to those preceding studies, we found a significant finite
volume effect between the results calculated on lattices with \((1.2\,
{\rm fm})^3\) and \((2.4\, {\rm fm})^3\) volumes \cite{Sasaki:2003jh}: about 20\% increase was observed.  On the large volume we find \(g_A = 1.212\pm 0.027({\rm stat})\pm 0.024({\rm
norm})\).  The quoted systematic error is the dominant known one,
corresponding to current renormalization.  This theoretical first
principles calculation, which does not yet include isospin breaking
effects, yields a value of \(g_A\) only a little bit below the
experimental one, \(1.2670 \pm 0.0030\) \cite{Hagiwara:2002fs}.  The
dependence on the quark mass on the larger volume is very mild, while
it is steep on the smaller.

\begin{figure}
\begin{center}
\includegraphics[width=70mm,bb=0 0 556 483]{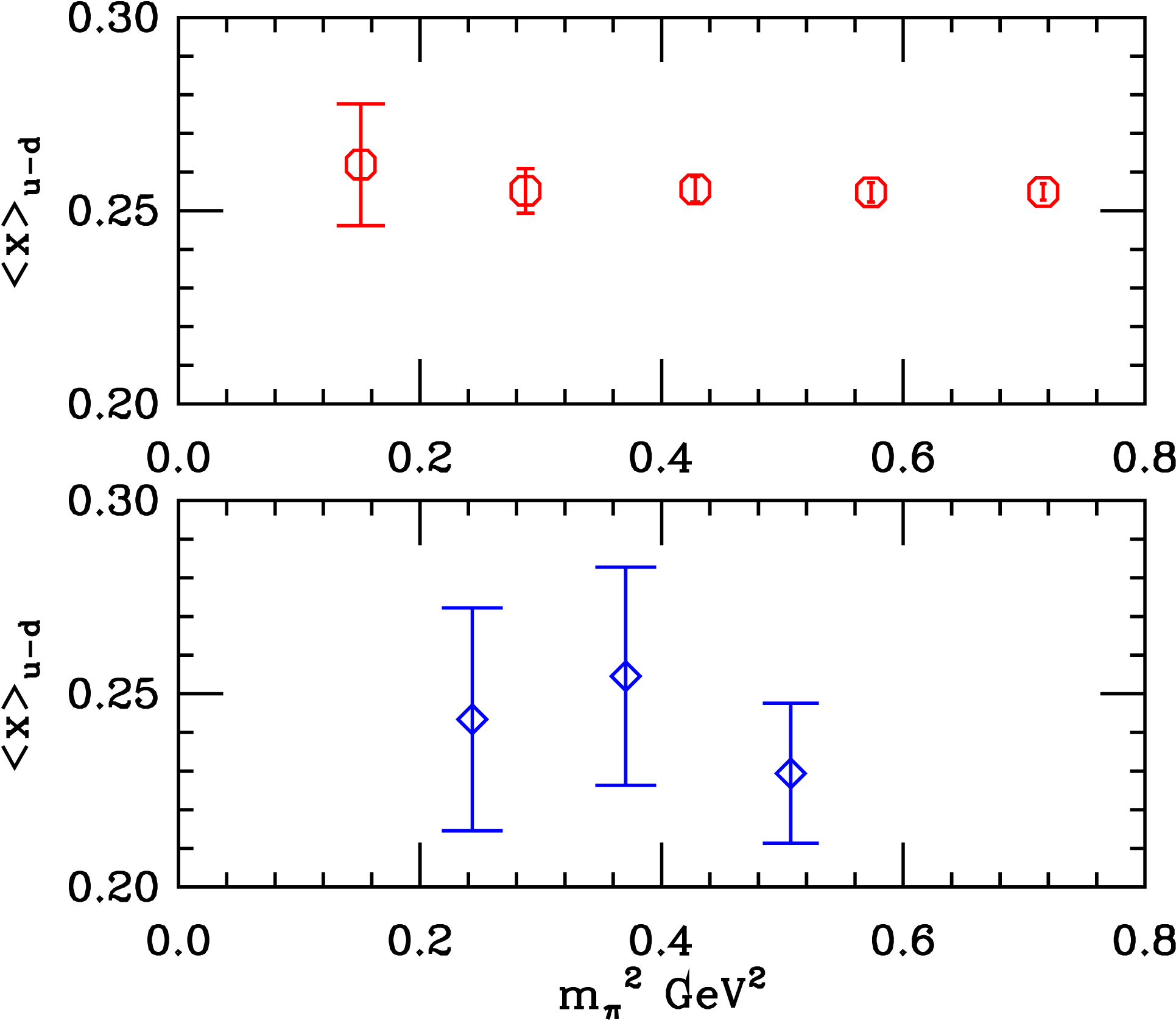}
\end{center}
\vspace{-1cm}
\caption{ The bare momentum fraction in
quenched (circle) and full-QCD (diamond).}
\label{fig:X}
\vspace{-.5cm}
\end{figure}
First few moments of nucleon structure functions are as easy as these
charges, in that they do not require finite momentum states.  The
structure functions such as \(F_1(x, Q^2)\), \(F_2(x, Q^2)\), \(g_1(x,
Q^2)\), and \(g_2(x, Q^2)\) are measured in lepton deep inelastic
scattering, and \(h_1(x, Q^2)\) in the RHIC Spin experiment.  Their
moments are factorized in the operator product expansion into Wilson
coefficients and non-perturbative quantities such as \(\langle
x^n\rangle_q\), \(\langle x^n\rangle_{\Delta q}\), \(d_n\) or
\(\langle 1 \rangle_{\delta q}\).  The latter are nucleon forward
matrix elements of certain local operators, such as \(\gamma D\),
\(\gamma_5 \gamma D\), and \(\gamma_5\sigma D\), and hence are
accessible by lattice QCD numerical methods similar to form factor
calculations.  There are calculations by QCDSF, QCDSF/UKQCD, and
LHPC-SESAM collaborations with (improved) Wilson fermions
\cite{Wilson}.  In Lattice 2002 we presented some preliminary quenched
results from the present DWF/DBW2 calculations \cite{Kostas02}.  The
DWF/DBW2 combination indeed works better: good chiral behavior, ease
of non-perturbative renormalization, and better approach to the
continuum were seen.  We present higher statistics quenched results
here, and some preliminary full-QCD ones.  We still lack
non-perturbative renormalization for some observables.

\begin{figure}
\begin{center}
\includegraphics[width=70mm,bb=0 0 546 484]{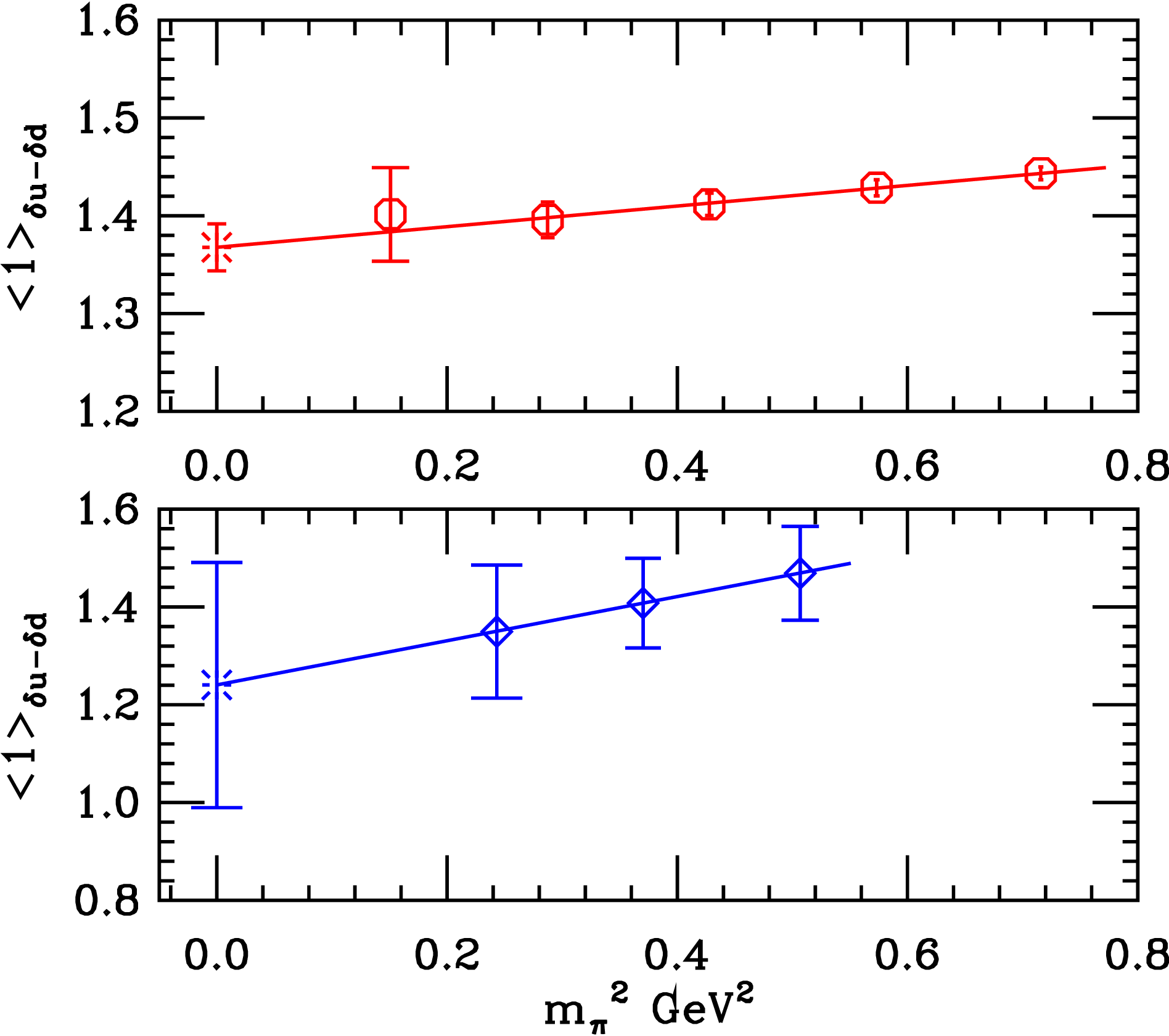}
\end{center}
\vspace{-1cm}
\caption{The bare \(\langle 1 \rangle_{\delta u - \delta d}\) matrix element in the quenched (circle) and full (diamond) QCD.}
\label{fig:h1}
\vspace{-.5cm}
\end{figure}
In the on-going full-QCD calculations, a \(16^3\times 32 \times 12\)
lattice is used with the DBW2 gauge coupling set at \(\beta\)=0.8 and
the domain-wall height at \(M_5\)=1.8.  Dynamical DWF quarks with mass
0.02, 0.03, and 0.04 (about strange mass) \cite{Chris03} in the
lattice units are used.  The nucleon structure calculations reported
here are done only at the dynamical quark mass, and no partially
quenched study has been made.  About 900--200 conjugate-gradient
iterations are necessary to achieve \(10^{-8}\) accuracy for the Dirac
equation solver.  Hybrid Monte Carlo acceptance of about 70-80 \% is
observed.  About 50 configurations are used at each dynamical quark
mass value.  The residual mass of \(m_{\rm res} = 0.00137\), \(\rho\)
meson mass \(m_\rho = 0.43(2)\) which translates into the lattice
cutoff estimate of \(a^{-1}\) \(\sim\) 1.8 GeV, and the nucleon mass
\(m_N = 0.56(4)\) or  \(\sim\)1.01(7) GeV are obtained.  Quality of
plateaus in the three-point functions is similar to the quenched case
at similar statistics.

\begin{figure}
\begin{center}
\includegraphics[width=70mm,bb=0 0 570 256]{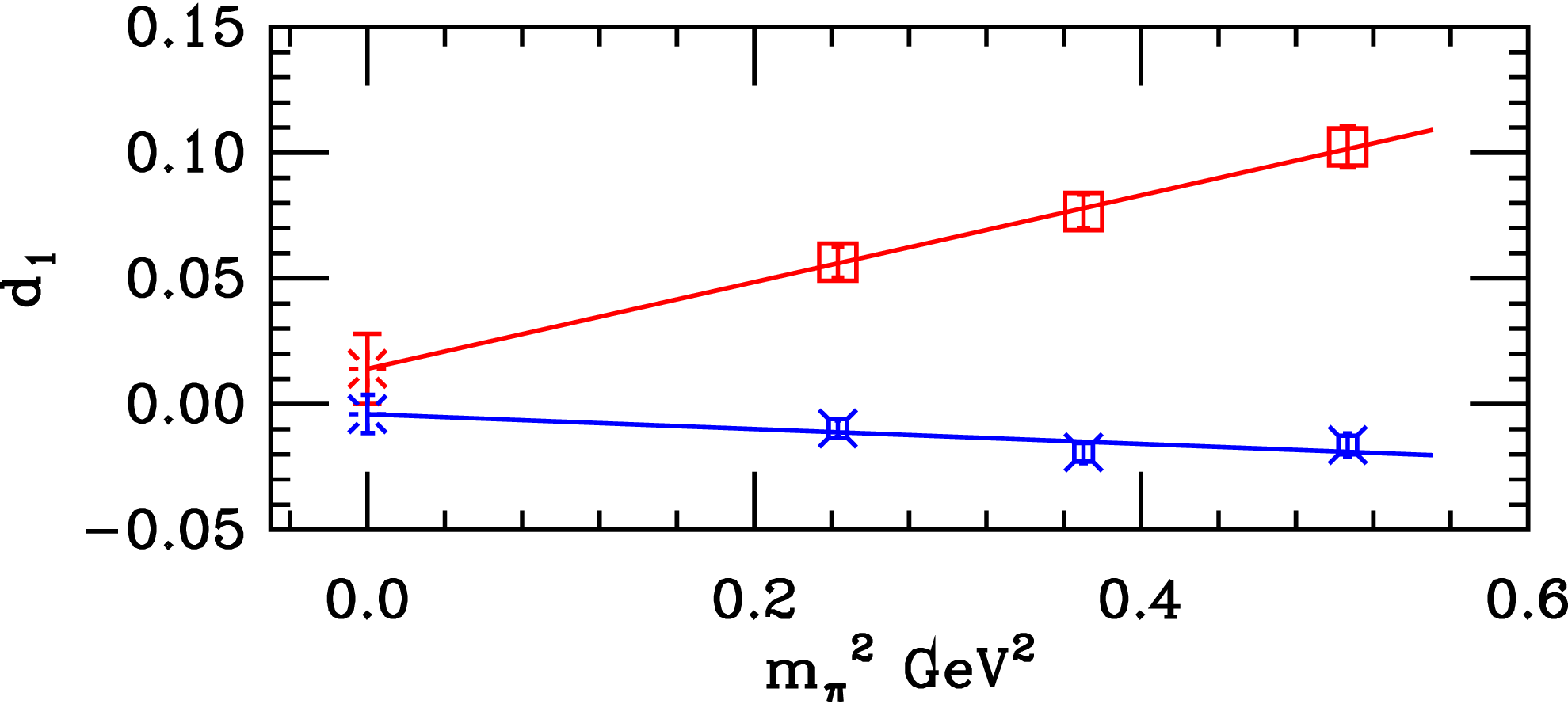}
\end{center}
\vspace{-1.cm}
\caption{The \(u\) (square) and \(d\) quark (cross) contributions to the \(d_1\) matrix element in full QCD.}
\label{fig:d1}
\vspace{-.5cm}
\end{figure}
In Fig.\ \ref{fig:mfdependence} we show the ratio of the isovector
axial and vector charges as a function of the pion mass squared.  The
full-QCD results (diamonds) are consistent with the old quenched
results (circles) which are also shown: no mass dependence is seen.
The weighted average of the three full-QCD points is \(g_{_A}/g_{_V}\)=1.21(8) and is consistent with the quenched results.

In Fig.\ \ref{fig:X} we present the quark momentum fraction in quenched (circle) and
full QCD (diamond).  Though not yet renormalized, in the quenched approximation it
does not show any curvature toward the chiral limit.  The first moment of the transversity is
shown in Fig.\ \ref{fig:h1}.  In both the quenched and full QCD we obtain the 
chiral limit of the bare matrix element by linear extrapolation.  We use the
RI-MOM scheme with one loop running to non-perturbatively renormalize the 
matrix element.  The quenched result is 1.194(25) (\(\overline{\rm MS}\), 2 GeV) and the full QCD result will be reported soon.

We have also calculated the \(d_1\) matrix element for which
lack of chiral symmetry results in a mixing with a lower dimensional
operator.  We reported last year that this mixing is suppressed with domain wall fermions  in the quenched approximation \cite{Kostas02}.  Our full-QCD \(d_1\) results, shown in Fig.\ \ref{fig:d1}, are again in  sharp contrast with the unsubtracted Wilson fermion results indicating that DWF suppress the power divergent mixing.

To summarize, the quenched calculations with DBW2 and DWF works well
for nucleon matrix elements.  The axial to vector charge ratio,
\(g_{_A}/g_{_V}\), is well reproduced to within 5 \% of the
experimental value even on a modest volume and relatively heavy quark
mass.  The dependence on quark mass is very mild.  Likewise the moments
of structure functions \(\langle x \rangle_q\), \(d_1\), and \(\langle
1 \rangle_{\delta q}\) all show mild linear dependence on quark mass,
though some of them are yet to be renormalized.  The full-QCD
calculations of these quantities seem to work as well.  Similar mild
linear dependence on the quark mass was observed, extrapolating to
estimates not contradicting with the quenched ones 
(where renormalization is done), albeit with
limited statistics.  With QCDOC computer \cite{QCDOC}, we will be adding more observables such as
flavor-singlet quantities and form factors and higher moments of
structure functions that require finite momentum, and moving to larger
lattice volumes and smaller quark masses.


\begin{thebibliography}{9}

\bibitem{DWF}
D.B.~Kaplan,
Phys.\ Lett.\ B {\bf 288}, 342 (1992); 
Y.~Shamir,
Nucl.\ Phys.\ B {\bf 406}, 90 (1993); 
R.~Narayanan and H.~Neuberger,
Phys.\ Lett.\ B {\bf 302}, 62 (1993); 
V.~Furman and Y.~Shamir,
Nucl.\ Phys.\ B {\bf 439}, 54 (1995).

\bibitem{RBC}
T.~Blum {\it et al.} [RBC], 
hep-lat/0007038, to appear in Phys.\ Rev.\ D;
Phys.\ Rev.\ D {\bf 66}, 014504 (2002); 
Phys.\ Rev.\ D {\bf 65}, 014504 (2002); 
RBRC Scientific Articles 4;
hep-lat/0110075, to appear in Phys.\ Rev.\ D.


\bibitem{Sasaki:2001nf}
S.~Sasaki {\it et al.},
Phys.  Rev.  D{\bf65}, 074503 (2002).  


\bibitem{DBW2}
T.~Takaishi,
Phys.\ Rev.\ D {\bf 54}, 1050 (1996);
P.~de Forcrand {\it et al.}  [QCD-TARO], 
Nucl.\ Phys.\ B {\bf 577}, 263 (2000);
Y.~Aoki {\it et al.},
hep-lat/0211023.


\bibitem{Wilson}
M.~Fukugita {\it et al.},
Phys.\ Rev.\ Lett.\  {\bf 75}, 2092 (1995); 
K.~F.~Liu {\it et al.},
Phys.\ Rev.\ D {\bf 49}, 4755 (1994); 
P.~Rakow  {\it et al.},
Phys.\ Rev.\ D {\bf 53}, 2317 (1996); 
S.~G{\" u}sken {\it et al.}  [TXL], 
Phys.\ Rev.\ D {\bf 59}, 114502 (1999);
D.~Dolgov {\it et al.}  [LHPC], 
Phys.\ Rev.\ D {\bf 66}, 034506 (2002), hep-lat/0201021;
S.~Capitani {\it et al.},
Nucl.\ Phys.\ Proc.\ Suppl.\  {\bf 79}, 548 (1999); 
R.~Horsley  [UKQCD], 
Nucl.\ Phys.\ Proc.\ Suppl.\  {\bf 94}, 307 (2001).  

\bibitem{Sasaki:2003jh}
S.~Sasaki {\it et al.},
hep-lat/0306007, to appear in Phys.\ Rev.\ D.

\bibitem{Hagiwara:2002fs}
K.~Hagiwara {\it et al.}  [Particle Data Group], 
Phys.\ Rev.\ D {\bf 66}, 010001 (2002).

\bibitem{Kostas02}
K.~Orginos [RBC], in Proc.  ``Lattice 2002,'' Nucl.\ Phys.\ B Proc.\ Suppl.\ 119, 386 (2003); in Proc.  ``Spin 2002,'' hep-lat/0211025.

\bibitem{Chris03}
C.  Dawson, in these proceedings.

\bibitem{QCDOC} 
T.  Wettig and K.  Petrov in these proceedings.

\end{thebibliography}
\end{document}